\documentclass[12pt,preprint]{aastex}
\shorttitle{$Fermi$ Discovery of NGC~1275}
\shortauthors{Abdo et al.}
\usepackage{times}
\begin{document}

\title{Fermi Discovery of Gamma-Ray Emission from NGC 1275}

\author{
A.~A.~Abdo\altaffilmark{1,2}, 
M.~Ackermann\altaffilmark{3}, 
M.~Ajello\altaffilmark{3}, 
K.~Asano\altaffilmark{4}, 
L.~Baldini\altaffilmark{5}, 
J.~Ballet\altaffilmark{6}, 
G.~Barbiellini\altaffilmark{7,8}, 
D.~Bastieri\altaffilmark{9,10}, 
B.~M.~Baughman\altaffilmark{11}, 
K.~Bechtol\altaffilmark{3}
% ...
}
\altaffiltext{1}{National Research Council Research Associate}
\altaffiltext{2}{Space Science Division, Naval Research Laboratory, Washington, DC 20375}
\altaffiltext{3}{W. W. Hansen Experimental Physics Laboratory, Kavli Institute for Particle Astrophysics and Cosmology, Department of Physics and SLAC National Accelerator Laboratory, Stanford University, Stanford, CA 94305}
\altaffiltext{4}{Interactive Research Center of Science, Tokyo Institute of Technology, Meguro City, Tokyo 152-8551, Japan}
\altaffiltext{5}{Istituto Nazionale di Fisica Nucleare, Sezione di Pisa, I-56127 Pisa, Italy}
\altaffiltext{6}{Laboratoire AIM, CEA-IRFU/CNRS/Universit\'e Paris Diderot, Service d'Astrophysique, CEA Saclay, 91191 Gif sur Yvette, France}
\altaffiltext{7}{Istituto Nazionale di Fisica Nucleare, Sezione di Trieste, I-34127 Trieste, Italy}
\altaffiltext{8}{Dipartimento di Fisica, Universit\`a di Trieste, I-34127 Trieste, Italy}
\altaffiltext{9}{Istituto Nazionale di Fisica Nucleare, Sezione di Padova, I-35131 Padova, Italy}
\altaffiltext{10}{Dipartimento di Fisica ``G. Galilei", Universit\`a di Padova, I-35131 Padova, Italy}
\altaffiltext{11}{Department of Physics, Center for Cosmology and Astro-Particle Physics, The Ohio State University, Columbus, OH 43210}
% ...

\begin{abstract}
We report the discovery of high-energy ($E>100$ MeV) $\gamma$-ray emission 
from NGC 1275, a giant elliptical galaxy 
lying at the center of the Perseus cluster of galaxies, 
based on observations made with the Large Area Telescope (LAT) of the $Fermi$ 
Gamma ray Space Telescope. 
 The positional center of the $\gamma$-ray 
source is only $\approx 3^{\prime}$  away from the NGC 1275 nucleus, 
well within the 95$\%$ LAT error circle of $\approx 5^{\prime}$.
The spatial distribution of $\gamma$-ray photons is consistent with a point source.  
The average flux and power-law photon index 
measured with the LAT 
from 2008 August 4 to 2008 December 5 are $F_{\gamma} = (2.10 \pm$
0.23) $\times$ $10^{-7}$ ph $(>100$ MeV) cm$^{-2}$ s$^{-1}$ and
$\Gamma$ = 2.17 $\pm$ 0.05, respectively. The measurements
are statistically consistent with constant flux during the four-month LAT observing period.
Previous EGRET observations gave an upper limit 
of $F_{\gamma} < 3.72\times 10 ^{-8}$ ph $(>100$ MeV) cm$^{-2}$ s$^{-1}$
to the $\gamma$-ray flux from NGC 1275.
This indicates that
the source is variable on timescales of years to decades, and therefore restricts the
fraction of emission that can be produced in extended regions of the galaxy cluster. 
Contemporaneous and historical radio observations
are also reported. The broadband spectrum of NGC
1275 is modeled with a simple one-zone synchrotron/synchrotron self-Compton model 
and a model with a decelerating jet flow. 
%The significant mismatch between EGRET and $Fermi$  
%may be related to historical activity of the nucleus 3C~84 observed 
%in both radio and optical wavelengths.

\end{abstract}


\section{Introduction}
\label{sec:intro}

The Perseus cluster\footnote{The Perseus cluster is  
Abell 426 at redshift $z$=0.0179 and luminosity distance $d_L$ = 75.3 Mpc,
for a Hubble constant $H_0 = 71$ km s$^{-1}$ Mpc$^{-1}$ in a flat universe with $\Omega_m = 0.27$, implying  
21.5 kpc/arcmin.} is the brightest cluster of galaxies 
in the X-ray band and has been the 
focus of extensive research over many 
years and wavelengths. The cluster hosts the giant elliptical galaxy 
NGC 1275 at its center. 
%According to the emission-line characteristics 
%of the optical spectrum, NGC 1275 is now classified as Sy 1.5 
%(Veron-Cetty \& Veron 1998) galaxy, but was thought to be a blazar 
%in the past due to its strong and rapid variability 
%(e.g., Angel \& Stockman 1980; see also Pronik, Merkulova \& Metik
%1999). 
NGC 1275 has been variously classified as a Seyfert 1.5 because of its
emission-line optical spectrum, where broad lines are detected
(Veron-Cetty \& Veron 1998), but also as a blazar due to the strong
and rapid variability of the continuum emission and its polarization
(e.g., Angel \& Stockman 1980; see also Pronik, Merkulova \& Metik
1999). In the radio, NGC 1275 hosts the exceptionally bright radio
source Perseus~A, also known as 3C~84.  The source 3C~84 has a strong, compact
nucleus which has been studied in detail with VLBI (Vermeulen et al
1994, Taylor et al. 1996, Walker et al. 2000, Asada et al. 2006).
These observations reveal a compact core and a bowshock-like southern
jet component moving steadily outwards at 0.3 mas/year (Kellermann et
al. 2004; Lister et al. 2009).  The northern
counterjet is also detected, though it is much less prominent due to
Doppler dimming, as well as to free-free absorption due to an intervening
disk. Walker, Romney \& Benson (1994) derive from these observations
that the jet has an intrinsic velocity of $0.3c-0.5c$ oriented at an
angle $\approx 30^{\circ}-55^{\circ}$ to the line of sight. 
Polarization has recently been detected in the southern jet 
(Taylor et al. 2006), suggesting
increasingly strong interactions of the jet with the surrounding environment.

The radio emission continues on larger scales, and shows a clear
interaction with the hot cluster gas.  Observations with $ROSAT$
(B\"{o}hringer et al. 1993) and later $Chandra$ (Fabian et al. 2003,
2006) reveal the presence of cavities in the gas, suggesting that the
jets of 3C~84 have blown multiple bubbles in the hot intracluster
medium.  Perseus is the nearest and best example of a prototypical
``cooling core" cluster in which the radiative cooling time of the
X-ray emitting gas is considerably shorter than the age of the
Universe.  For a $\beta$-model (Cavaliere \& Fusco-Femiano 1976), 
the core radius of the Perseus cluster is $r_{\rm c}$ $\sim$ 0.3 Mpc (or
$\sim$ 0.3$^{\circ}$; see,  e.g., Ettori et al. 1998; Churazov et
al. 2003). Heating by the central AGN is thought to be
responsible for balancing the radiative cooling, although the exact
mechanisms by which the energy is transported and dissipated are still 
unclear.  Shocks and ripples are clearly evident in the deep $Chandra$
image of Perseus, and could provide steady heating of the center of
the cluster (Fabian et al. 2005; 2006).  On even larger scales,
Perseus is one of the few clusters exhibiting a mini-halo of size
$\sim$300 kpc seen in low-frequency radio emission (Burns et
al. 1990).  This mini-halo is presumed to arise from synchrotron
emission from widely distributed relativistic particles and fields
energized in the central regions of the cluster.

Furthermore, the Perseus cluster appears to contain 
a nonthermal component, namely an excess of hard X-ray 
emission above the thermal bremsstrahlung from the diffuse 
hot cluster gas. Based on a deep $Chandra$ observation, the 
non-thermal X-ray component has been mapped over the core 
of the cluster and shows a morphology similar to the radio mini-halo 
(Sanders, Fabian \& Dunn 2005; Sanders \& Fabian 2007). 
This claim was, however, questioned 
%by other group who employed a 
on the basis of a long $XMM$-$Newton$ exposure (Molendi \& Gastaldello 2009). 
Above 10 keV, a hard X-ray component has been detected with 
$HEAO$-$1$ (Primini et al. 1981) and $BeppoSAX$/PDS (Nevalainen et
al. 2004), although it was not detected with $CGRO$/OSSE in the 
0.05$-$ 10 MeV range (Osako et al. 1994). 
More recently, 
%$Swift$/BAT reports about the 
%detection of 10 cluster of galaxies in the 15$-$55 keV range 
10 galaxy clusters were detected in the 15$-$55 keV range with $Swift$/BAT
(Ajello et al. 2009). 
Perseus is the only cluster that displays a high-energy 
non-thermal component up to 200 keV, but the hard tail 
seen in the BAT spectrum is likely due to nuclear emission from NGC 1275 
%and not to a non-thermal component originated in the 
rather than to non-thermal emission from the intercluster medium. This idea 
is supported by possible flux variations compared to past hard X-ray 
observations, and by the fact that the extrapolation of the BAT spectrum 
is in good agreement with the luminosity of the nucleus as measured with 
$XMM$-$Newton$ (Churazov et al.\ 2003).

At higher energies, $\gamma$-ray observations toward 
NGC 1275 and the Perseus clusters were first reported in the 1980's by 
Strong \& Bignami (1983). The $COS$ $B$ data, taken between 
1975$-$1979 (Strong et al. 1982; 
Mayer-Hasselwander et al.\ 1982), show a $\gamma$-ray 
excess at the position of the galaxy, although evidence for 
emission $uniquely$ related to NGC 1275 is ambiguous 
(positional uncertainties were not given for the $COS$ $B$ data). 
Interpreted as emission from NGC 1275, the $\gamma$-ray flux was 
$F_{\gamma}$ = 8.3 $\times$ $10^{-7}$ ph($>70$ MeV) cm$^{-2}$ s$^{-1}$. 
Further observations in the MeV-GeV range 
were made by $CGRO$/EGRET in the 1990's as part of a
search for $\gamma$-ray emission from 58 clusters of galaxies 
between 1991 and 2000 (Reimer et al. 2003). No evidence 
was found for high-energy $\gamma$-ray emission of individual 
clusters, nor as a population. The 2$\sigma$ 
upper limit for the Perseus cluster/NGC~1275 is 
$F_{\gamma}$ $<$ 3.72$\times$$10^{-8}$ ph ($>100$ MeV) cm$^{-2}$ s$^{-1}$, 
which is more than an order of magnitude 
lower than the flux reported by $COS$ $B$. 
Observations with improved sensitivity, now possible with $Fermi$, 
are crucial to  
confirm $\gamma$-ray emission from NGC 1275 and possible time variability.

There are several reasons to think that the Perseus/NGC~1275 (3C~84)
system could be a $\gamma$-ray emitter.  First, a few extragalactic
$non$-blazar sources, namely Centaurus A, an FR I radio galaxy (Sreekumar et al. 1999),
3C~111, a broad line radio galaxy (Nandikotkur et al.\ 2007; Hartman et
al. 2008), and possibly the radio galaxy NGC~6251 (Mukherjee et al. 2002), were already
detected with EGRET. In contrast to blazars, which form the majority
of extragalactic $\gamma$-ray sources (Hartman et al.\ 1999), most
radio galaxies have large inclination angles and hence there is no
significant amplification of the emission due to Doppler beaming. 
However, if the jet has velocity gradients (see  Georganopoulos \&
Kazanas 2003b for a decelerating flow and Ghisellini et al. 2005 for a
spine-sheath velocity profile), as suggested by recent radio/X-ray 
observations (e.g., for transversal profiles, see 
Laing \& Bridle 2002; Kataoka et al. 2006), 
it is possible to produce bright $\gamma$-ray emission from the nuclei
 of some radio galaxies via the inverse Compton process where the
 emission from the slow part is amplified in the rest-frame of the 
faster part, and vice-versa. 
Second, $\gamma$-ray emission from the
cluster is also expected as a result of (1)
secondary nuclear interactions of high-energy cosmic rays with the intercluster
medium or as the origin of a secondary population of relativistic
electrons (Berezinsky et al.\ 1997, Atoyan \& V{\"o}lk 2000); (2)
particle acceleration at large-scale scale shocks in forming clusters
(e.g., Totani \& Kitayama 2000), or at a shock excited by an AGN
outburst at the cluster center (Fujita et al. 2007); and (3) dark
matter annihilation, which also acts as a heat source in the core of
cooling flow clusters (Totani 2004).  In contrast to the emission from
a compact AGN region, $\gamma$-rays from clusters would be steady 
on the observing time scales.  Hence time variability, if
detected, provides an important clue to the origin of the 
$\gamma$-ray emission.

With the successful launch of $Fermi$ (formerly known as $GLAST$), we
have a new opportunity to study
$\gamma$-ray emission from radio galaxies and cluster of galaxies with
much improved sensitivity. As a  first step, we report here the
$Fermi$ discovery 
of $\gamma$-ray emission from NGC 1275.  In $\S$2, we describe the
$Fermi$ $\gamma$-ray observations, 
data reduction process, and analysis results.
In $\S$3 we present historical radio and contemporaneous 
radio observations with the UMRAO, RATAN and the MOJAVE program.
Our results are presented in the context of jet emission models in \S
4.  Conclusions are given in  $\S$5.

\section{$\gamma$-Ray Observations }
\label{sec:observations}

On June 11, 2008, the $Fermi$ Gamma-ray Space Telescope was
successfully launched into a low-Earth orbit at $\sim 565$ km, with an inclination angle
of 25.5$^{\circ}$.  The LAT instrument on $Fermi$
is described in detail in Atwood et al.\ (2009) and references
therein. The LAT relies on the conversion of $\gamma$ rays
into electron-positron pairs; tracking of those pairs allows the
determination of the direction of the incident $\gamma$-ray. Such a
design results in a wide field of view ($\simeq$ 2.4 sr),
simultaneously available to the detector.  Compared to earlier
$\gamma$-ray missions, the LAT has a large effective area ($\sim$
8,000 cm$^2$ on axis at 1 GeV for the event class considered here), 
wide energy coverage from $\approx 20$ MeV to
$>300$ GeV, improved angular resolution (a point-spread function of 
$\sim$ 0.6$^{\circ}$ at 1 GeV for 68$\%$
containment) and is live about 90$\%$ of the time. 

During the first year of operations, most of the telescope's time is
being dedicated to ``survey mode'' observing, where $Fermi$ points 
away from the Earth, and nominally rocks the spacecraft axis 
north and south from the orbital plane to enable monitoring of the 
entire sky on a time scale shorter than a 
day or less.
The whole sky is surveyed every 
$\sim$3 hours (or 2 orbits). The first light images of the 
$\gamma$-ray sky are found in the LAT official web page.\footnote{http://www.nasa.gov/mission\_pages/GLAST/main/index.html}
We report here on the LAT's initial observations of the Perseus/NGC 1275 
region, using data collected during the first four months of the on-going 
all-sky survey. 
The source was first detected during the Launch and 
Early Operation phase (L\&EO, namely
the period lasting approximately 60 days after the launch until August 3).
However, because the instrument configuration was not tuned for optimum 
performance, we concentrate our analysis on the survey data starting from 
2008 August 4.

\subsection{Data Reduction}

The data used here comprises all scientific data obtained between 
4 August 2008 and 5 December 2008. The interval runs from MET 239557417 to 
250134308. We have applied the zenith angle cut to eliminate photons 
from the Earth's limb, at 105$^{\circ}$. 
This is important 
in pointed mode observations, but also important 
for survey mode due to overshoots and sun avoidance maneuvers.
 The same zenith cut 
is also accounted for in exposure calculation using the LAT 
science tool \textsc{gtltcube}.
We use the ``Diffuse'' class events (Atwood et al.\ 2009), 
which are those reconstructed 
events having the highest probability of being photons.

In the analysis presented here, we set the lower energy bound 
to  a value of 200 MeV, since the bin count for photons with 
energies of $\approx 100$ 
MeV and lower is systematically lower than expected 
based on extrapolation of a reasonable function. Note that theta cuts, 
which would eliminate events close to the edge of the field of view, are not
applied in the present analysis since we still need to study the tradeoffs 
introduced by the cut versus those introduced by not having the cut.
Science Tools version v9r8p3 and IRFs (Instrumental Response Functions) 
\textsc{P6\_V1} (a model of the spatial distribution of photon events
calibrated pre-launch) were used throughout this paper. 

\subsection{$\gamma$-Ray Results}
\label{sec:results}

Figure 1 shows a close-up of the $Fermi$ image above 200 MeV centered on the
position of NGC 1275 (RA = 49.951$^\circ$, DEC = 41.512$^\circ$), with
an image radius of the Region of Interest (ROI) $r$ = 8 deg. The image has been smoothed with a
two-dimensional Gaussian function with $\sigma$ = 0.2$^\circ$. 
The extended feature towards the upper left is the edge of the Galactic diffuse 
emission. The brightest source is located at right ascension $= 50.000^\circ$, declination $=41.524^\circ$, 
and coincides within the uncertainties with the direction to NGC 1275.  
The positional center of the $\gamma$-ray emission is only 0.05$^\circ$ 
from the position of the NGC 1275 nucleus, well within the 95\% LAT 
error circle of 0.086$^\circ.$\footnote{More accurately, we should call
this the ``NGC~1275/Perseus region,'' since at this stage it is still unclear
  whether the $\gamma$-ray emission comes from the nucleus of NGC~1275
  or the Perseus cluster. This will be discussed later in the discussion
  section $\S$4. Also there are several galaxies, NGC 1273, 1274,
  1277, 1278, and 1279 in the LAT error circle, but NGC 1275 is by far
  the brightest, is strong in the radio, and is the closest source to
  the $\gamma$-ray peak position.}

%The second brightest source in the west is most likely the high-frequency peaked BL Lac 3C~66A, which is 0.014$^{\circ}$ apart
%and well within the LAT error circle of 0.054$^{\circ}$ ($ref$: 
%Tosti et al 2008; also AGN population paper). 
%Also this is possibly associated with the EGRET 
%source 3EG J0222+4253, and TeV $\gamma$-ray emission has recently 
%been reported by the MAGIC collaboration (Aliu et al. submitted).  

%In addition, there are number of 
%$\gamma$-ray sources whose positions and fluxes  will be published 
%in a forthcoming catalog paper (Abdo et al. 2009a; submitted) and 
%bright AGN source list from the first three months of the $Fermi$ LAT 
%all sky survey: Abdo et al. 2009b, submitted).

Figure 2 shows the projection of the $\gamma$-ray images in low (0.2$-$1 GeV) 
and high (1$-$10 GeV) energy bands, specifically, sliced photon count 
distributions projected onto a RA axis with $\Delta$$\theta_{\rm DEC}$ 
= $\pm$ 2$^{\circ}$, centered on NGC~1275 (black points with errors; 
bin width 0.1 $^{\circ}$). The most prominent peak in the center is NGC~1275, 
while the smaller peak is also seen in the east (``src\_A''). 
Red solid lines show the best-fit model determined from the 
likelihood analysis described below, in which we assume two point 
sources (NGC~1275 and src\_A) with the Galactic and extragalactic 
diffuse emission overlaid. 
The LAT has an angular resolution of $\theta_{68}\simeq$ 0.6$^{\circ}E_{\rm
GeV}^{-0.75}$ (Atwood et al. 2009), giving $\theta_{68}\approx 2.0^{\circ}$ 
at 200 MeV. The counts distributions of NGC~1275 and src\_A are 
 consistent with this distribution in low- and high-energy bands, 
indicating that the diffuse extended component combined with a point source 
for src\_A does not contaminate NGC~1275/Perseus, at least within current 
photon statistics.

To study the average spectrum of NGC~1275 during the four-month
observation, we use the standard
maximum-likelihood spectral estimator provided with the LAT science
tools \textsc{gtlike}.  This fits the data to a source model, along with models
for the uniform extragalactic and structured Galactic backgrounds.  As
shown in Figure 1, the upper left count map ($\simeq$
10$^{\circ}$ from the NGC 1275 nucleus) is dominated by the bright soft
$\gamma$-ray emission of the Galactic plane. Since the distribution
and amount of the Galactic diffuse emission itself is still a matter
of debate, careful choice of the source region is important especially
for relatively faint sources. We therefore made several trials by
changing the ROI radius from 5$^{\circ}$ to
20$^{\circ}$ in steps of 5, 8, 10, 12, 15, 20$^{\circ}$, 
respectively.\footnote{The LAT team recommends that the ROI should 
be at least 15$-$20$^{\circ}$ in confused regions near the Galactic 
plane, $\le$ 10$^{\circ}$ for isolated high latitude regions.}
We used a recent Galactic diffuse model, 54\_59Xvarh7S, 
generated using GALPROP with the normalization free to 
vary in the fit. The response function used 
is \textsc{P6\_V1\_DIFFUSE}.

Since a different choice of ROI yielded essentially the same results
within statistical uncertainties, we set $r$ = 8$^{\circ}$ in the
following analysis to minimize contamination from the Galactic plane
and nearby sources (for details, see the $Fermi$ LAT 
bright $\gamma$-ray source list; Abdo et al. 2009)  
and to reduce computational time; this region is 
large enough to contain most of the photons even at the lowest
energies where the LAT PSF broadens. 
With this choice, the only sources to be included in the modeling 
are NGC 1275, src\_A, and the Galactic and 
extragalactic emission as underlying diffuse background components.
We have also checked the contribution from sources outside the ROI, but
found it to be completely negligible. 

We model the continuum emission from both NGC 1275 and src\_A with a
single power law. The extragalactic background is assumed to have a
power-law spectrum, with its spectral index and the normalization
free to vary in the fit.  
From an unbinned \textsc{gtlike} fit the best fit power-law parameters
for NGC~1275 are:
\begin{equation}
\frac{dN}{dE} = (2.45 \pm 0.26)\times10^{-9} \;(\frac{E}{100 {\rm
 MeV}})^{-2.17\pm0.04}
 \hspace{5mm}{\rm ph\hspace{2mm}cm^{-2} ~s^{-1} ~MeV^{-1}},
\end{equation}
or 
\begin{equation}
F_{\gamma}~ = (2.10~ \pm ~0.23)~ \times~ 10^{-7}
 \hspace{5mm}{\rm ph(>100 MeV) ~cm^{-2} s^{-1}}, 
\end{equation}
\noindent where only statistical errors are taken into account and the 
spectrum was extrapolated down to 100 MeV. Systematic
errors for the LAT are still under investigation (Bruel et al.\ 2009, in
prep), but for relatively faint source like NGC~1275, the uncertainty
is dominated by statistical errors.

The predicted photon counts from NGC~1275 in the ROI are $N_{\rm
  pred}$ = 866.5 and test statistic (defined as TS = 2(log $L$ - log
$L_0$), where $L$ and $L_0$ are the likelihood when the source is
included or not) is $TS$ = 1206.6 above 200 MeV, corresponding to a 35
$\sigma$ detection. For the Galactic diffuse background, the
normalization is 1.050$\pm$0.026 and $N_{\rm pred}$ = 11542.8. The
near unity normalization suggests that the Galactic diffuse emission
estimated in the ROI is in good agreement with the current GALPROP
model.  The power-law photon index of the extragalactic background is
$\Gamma$ = 2.14$\pm$0.04 with $N_{\rm pred}$ = 2793.2.  This
spectral shape is consistent with what has been measured
with $CGRO$/EGRET ($\Gamma$ = 2.10$\pm$0.03) but the normalization
determined with $Fermi$, (1.23$\pm$0.16)$\times$10$^{-7}$ ph
  cm$^{-2}$ s$^{-1}$ MeV$^{-1}$ when extrapolated to 100 MeV, 
is about 30 $\%$ lower than that measured with 
EGRET (Sreekumar et al. 1998).  Although we
have considered src\_A, the source turned out to be weak and did not 
affect the analysis results presented here\footnote{In summary, the
  best-fit parameters for a power-law function gives the photon index 
$\Gamma$ = 1.92$\pm$0.20 with $F_{\gamma}~ = (0.25~ \pm ~0.16)~ \times~
  10^{-7} {\rm ph(>100 MeV) ~cm^{-2} s^{-1}}$ for src\_A.}. 
Figure 3 shows the LAT spectrum of NGC~1275 obtained by 
separately running \textsc{gtlike} for seven energy bands; 
200$-$400 MeV, 400$-$800 MeV, 800 MeV$-$1.6 GeV, 1.6$-$3.2 GeV, 
3.2$-$6.4 GeV, 6.4$-$12.8 GeV, and 12.8$-$25.6 GeV, 
where the dotted line shows the best fit power-law function for the 
NGC 1275 data given by Equation (1). 

Finally, we investigate the flux variations of NGC~1275 from 
August 4 to December 5 in 2008. To this end, we accumulated spectra 
with a time resolution of 7 days and fit each spectrum with the same
model as above. The ROI radius ($r$ = 8$^{\circ}$), energy range ($E$ $>$ 200
 MeV), and other screening conditions were the same as 
described  above. Since variability is $not$ expected for underlying 
background diffuse emission, we fixed the best-fit parameters 
as to an average values determined from the four-month integrated
spectrum for the Galactic/extragalactic background components. 
In this manner, only four parameters (power-law photon indices and 
normalizations for both NGC~1275 and src\_A) are set to be free for
the time-resolved spectral fits. 

Figure 4 shows the plot of the flux ($E$ $>$100 MeV: $upper$) and
photon index ($lower$) versus time.  It appears that the
flux of NGC 1275 may decrease gradually over four months, and there are
some hints of spectral evolution as well. However, 
%it is still premature to confirm this apparent
%variability since the photon statistics are still poor to reject
%constancy of the fitting parameters ($\chi^2$ = 12.2 and 12.4 for 16
the hypothesis of constancy cannot be rejected, with $\chi^2$ = 12.2
and 12.4 for 16 degrees of freedom, for flux and photon index
variations, respectively\footnote{NGC 1275 is flagged as a variable 
source in the $Fermi$ LAT bright $\gamma$-ray source list (Table. 6 
of Abdo et al. 2009). This is because they have fixed the spectral 
index of each source to the best fit value over the full interval 
to avoid large error bars in the flux estimates, whilst both flux and photon 
index free to vary in the fit of this paper. Further long-term
monitoring is thus important to confirm the variability of this source.}.  
We checked that the contaminant src\_A does not vary. 
%Similar to the spectral analysis,
We independently checked the light curve using \textsc{gtexposure},
taking a small ROI radius of $r$ = 2$^{\circ}$ to reduce 
the contamination from diffuse background and nearby sources.
We assumed the spectral photon index of $\Gamma$ = 2.2, and background 
was subtracted from nearby region of the same ROI radius. 
The results were consistent with what has been obtained 
with \textsc{gtlike}.
Further long-term monitoring of this source is important.
Since the source is apparently variable on longer time scale, 
year-scale variability is naturally expected as we will discuss 
below.

\section{Radio Observations}

In the radio, the University of Michigan Radio Astronomy Observatory
(UMRAO) have monitored 3C~84  
since 1965. The UMRAO
variability program utilizes a 26-meter prime focus paraboloid 
equipped with transistor-based radiometers which operate at the central
frequencies  4.8, 8.0, and 14.5 GHz; the bandwidths are 560, 760, and
1600 MHz respectively.  A typical observation consists of 8 to 16 
individual measurements obtained over a 20 to 40 minute time interval. 
The flux scale is set by observations of Cassiopeia A 
(e.g., see Baars et al. 1977). Further details of the 
UMRAO calibration and data analysis procedures are given 
in Aller et al. (1985). Figure 5 shows a long-term light curve of 
3C~84 measured at 14.5 GHz, taken by the UMRAO from 
February 1974 to December 2008. 
Interestingly, the radio flux density reached a maximum between 1980 and 1985 
(the $COS$ $B$ era), and then substantially faded out after 1990 
(O'Dea et al. 1984; Ter{\"a}sranta et al. 2004) when EGRET was observing.  
This trend appears  similar to the 
optical activities of this source (Nesterov et al. 1995; 
Pronik et al. 1999). 
Furthermore, the UMRAO light curve shows a flare (or a rising state) 
starting in 2005, which could be interpreted as an
ejection of new jet components. 

In fact, the MOJAVE (Monitoring Of Jets in Active galactic nuclei with
VLBA Experiments; Lister et al. 2009) 15~GHz VLBA observations of
3C\,84, taken simultaneously with {\it Fermi} on 2008~August~25, show a
significant brightening of the central sub-parsec-scale structure,
indicating that a flare is happening in the innermost jet region
(Figure~6). This brightening might be connected to the $\gamma$-ray
activity detected.
The 1--22~GHz instantaneous radio 
spectrum of 3C\,84 was also observed with the 600-meter ring radio 
telescope RATAN-600 of the Special Astrophysical Observatory, 
Russian Academy of Sciences, located in 
Zelenchukskaya, Russia, on 2008~September 11 and 12. The continuum 
spectrum was measured on both days quasi-simultaneously (within several
minutes) in a transit mode at six different bands with the following
central frequencies (and frequency bandwidths): 0.95~GHz (0.03~GHz),
2.3~GHz  (0.25~GHz), 4.8~GHz  (0.6~GHz), 7.7~GHz  (1.0~GHz),
11.2~GHz (1.4~GHz), 21.7~GHz (2.5~GHz). Details on the method 
of observation, data processing, and amplitude calibration are 
described in Kovalev et al. (1999). 
An average spectrum is used for the spectral energy 
distribution (SED).

\section{Discussion and Interpretation}
\label{sec:discussion}

In the previous sections, we have reported  the detection of
$\gamma$-ray emission from NGC~1275 during the initial
sky survey with $Fermi$, and historical and contemporaneous
radio observations with UMRAO, RATAN, and MOJAVE. Although excess $\gamma$-ray emission
around the position of this galaxy had been previously found with $COS$
$B$, the association of the latter with NGC~1275 was ambiguous, due to
the relatively poor angular resolution and low photon statistics
(Strong \& Bignami 1983; see $\S$ 1). The $Fermi$ observations, with
much improved sensitivity and angular resolution, allow us to more precisely determine the
localization of the $\gamma$-ray source and its possible association with NGC~1275.
%Even if the $\gamma$-ray emission first reported by $COS$ $B$ is
%actually related with NGC~1275, more serious problem is that the
More intriguing is that the source was not detected during 
 $CGRO$/EGRET observations over ten viewing periods (Reimer et al.\ 2003).  The
2$\sigma$ EGRET upper limit to the flux is $F_{\gamma}$ $<$ 3.72 $\times$10$^{-8}$
ph($>100$ MeV) cm$^{-2}$ s$^{-1}$, which is about factor of seven
$lower$ than the flux measured by $Fermi$/LAT, and
more than an order of magnitude lower than the $COS$ $B$ flux (see Fig.\ 7).
This means the source varies on timescales shorter than years to
decades, so that the emission 
region size $R\lesssim c t_{\rm var} \approx 0.3$ pc.

With this simple estimate, we can provide 
useful constraints on whether the $\gamma$-ray emission originates 
from a cluster or AGN.  
Although the LAT error circle is  still large enough to 
%mix the non-thermal AGN with thermal cluster
include both non-thermal AGN and nonthermal cluster 
emission, a large fraction of the $\gamma$-ray emission measured with the
$Fermi$ LAT must  originate from within a few light-years of an active region, 
most likely the cluster center, 
on the basis of the EGRET upper limit.
%must be most inner site of the cluster (i.e., very close to the nucleus of NGC~1275; 
%see $\S$ 3) 
Since the Perseus cluster is extended over $\gtrsim$ 0.5$^{\circ}$
(or $\beta$ radius $\sim$ 0.3$^{\circ}$; see $\S$ 1), corresponding to hundreds of kpc,
if the 
emission were extended on this size scale, it would not have been variable and 
could have been detected as an extended source with the LAT above 1 GeV,
where the PSF becomes smaller than $\approx 0.5^\circ$. As seen in Figure 2, however, the
observed count distribution is consistent with a point source.  

This limits the $\gamma$-ray flux from the cluster formed by (1) p-p
interactions of high-energy cosmic rays or by (2) particle acceleration
at a large-scale shock to the flux upper limit measured with EGRET.
(see $\S$ 1). Thus, the bulk of the $Fermi$ emission is limited to 
a region of a few light years in extent.
%Note that even relatively
%small, arcmin-scale structures correspond to a few tens of kpc (21.5
%kpc/arcmin) at the distance of the Perseus Cluster ($z$ = 0.0179) and
%hence flux variability as fast as years is not expected.  
One may also suspect that the high energy $\gamma$-ray emission 
could be related to the
``cavities" seen in the X-ray images of the Perseus Cluster (Fabian et
al. 2000; 2003), which are likely inflated by the jet
from 3C~84. Their size scale, on the order of arcminutes, is too
large to account for the inferred time variability.

Another possibility is the $\gamma$-ray flux originates from 
the annihilation of dark matter particles, for example, neutralino
dark matter.  An annihilation $\gamma$-ray signal from
the whole cluster would be extended and inconsistent with the
$Fermi$ observation.  In any case, the expected flux is much smaller when a
standard annihilation cross section is assumed.  However, the growth
of the supermassive black hole at the cluster center may produce a
spike in the density profile, resulting in a much higher annihilation
rate from the central region within $\sim$ 0.1 pc (e.g., Totani 2004; Colafrancesco et al.\ 2006). This
annihilation emission should be observed as a point source for the LAT
resolution, and flux modulation on the dynamical time scale ($\sim$ 4 
months within 0.1 pc) is possible.  However, the continuum gamma-ray
spectrum from neutralino annihilation should be strongly peaked at
$\sim$1$-$10 GeV in the standard framework of particle physics, which
conflicts with the observed LAT spectrum (Figure 3 and Figure 7). 

The available evidence appears to be most consistent with $\gamma$-ray emission 
arising from the pc-scale AGN jet. Nonthermal nuclear emission 
is also detected at other wavelengths. 
Recent $Chandra$ and $XMM$-$Newton$ observations with
excellent angular resolution to resolve the nucleus revealed that
non-thermal nucleus emission is well represented by a simple power-law
function of $\Gamma$ = 1.65 (Churazov et al. 2003; Balmaverde et
al. 2006; Molendi \& Gastaldello 2009), with some hints of flux
variations. Also, hard X-ray emission detected with $Swift$/BAT is
likely due to non-thermal emission from the nucleus of NGC~1275
(Ajello et al. 2009).  In the optical, nuclear variability of NGC~1275 
has been indicated in densely monitored observations since the 1960's
(Nesterov et al. 1995; Pronik et al. 1999). The source was highly
variable and bright in the 1970's ($m_R$ $\simeq$ 12.0$\pm$0.5 mag),
suddenly faded in the 1980-90's with $m_R$ $\simeq$ 13.5 mag
(Ciprini et al. 2008), and has been gradually rising up again after 2000. 
To gauge the recent optical activity of NGC~1275, we measured the source flux
in six filter Swift/UVOT (Roming et al. 2005) images from Dec 2007 (Fig.~7) using $r=5''$ circular regions.
Additionally, our recent optical observations by MITSuME (Multicolor Imaging Telescope for 
Survey and monstrous Explosions; e.g., Kotani et al. 2005)
exhibits preliminary results of $m_R$ $\simeq$
12.7$\pm$0.3 mag during the $Fermi$/LAT observation in June-Sept 2008.
Interestingly, it appears that the optical flux traces the historical
$\gamma$-ray activity from $COS$ $B$ to the $Fermi$ era. Note, however, 
that care must 
be taken when comparing results from different telescopes due to host galaxy
subtraction and different techniques to calculate photon counts (Nilsson
et al. 2007).

Figure 7 shows the overall $\nu F_\nu$ SED of NGC 1275 constructed 
with radio to $\gamma$-ray multiband data.
Although the archival NED (NASA/IPAC extragalactic database) data 
contain host galaxies contamination in optical, the non-thermal nuclear 
spectrum shows two pronounced continuum components, one  peaking between the 
optical and IR and the other in the $\gamma$-ray regime. 
In analogy with blazars (e.g., Kubo et al. 1998; Fossati et al. 1998), 
the low energy component is probably due to synchrotron radiation of 
relativistic electrons accelerated within the outflow, while Compton
scattering by the same electron is most likely responsible for 
the nonthermal X-ray and  high energy $\gamma$-ray component. As can be seen, 
the $\gamma$-ray flux is comparable in apparent luminosity with the lower 
energy radio/optical 
flux. Hence the overall SED appears to be similar to low 
frequency peaked BL Lac objects (e.g., Kubo et al. 1998) in that 
the low energy peak is in the IR-optical and the high peak is 
in the soft gamma-rays.  Two low peaked BL Lacs have been observed 
with TeV gamma-rays:  BL Lac (Albert et al. 2007) and 3C 66A 
(Acciari et al. 2009) and thus NGC 1275 is a potential TeV source as well.

We use two models to fit the SED of the nonthermal emission of NGC 1275
in Figure 7. First we consider a simple one-zone SSC model fit to the 
$Fermi$ data and contemporaneous radio data (blue dashed curve; 
see Finke et al.\ 2008 for details).
This model employs a jetted outflow with bulk Lorentz factor $\Gamma =
1.8$ and Doppler factor $\delta = 2.3$, so that the observing angle to
the jet direction is $\theta = 25^\circ$. 
The mean magnetic field in the radiating plasma is $B = 0.05$ G, 
and the comoving radius of the jet emission region is 
2$\times 10^{18}$ cm, corresponding to a variability timescale of 
$\approx 1$ yr. The nonthermal electron distribution 
is assumed to be described by a broken power law with number indices 
(where the electron distribution $n(\gamma) \propto \gamma^{-p}$)  
$p_1 = 2.1$ for $800 \lesssim \gamma \leq 960$, and index $p_2 = 3.1$ 
for $960 \leq \gamma \leq 4 \times 10^5$, where $\gamma$ is the electron 
Lorentz factor in the fluid frame, and the Poynting flux density is
about twice the electron energy density. This simple homogeneous model 
provides an adequate fit the NGC~1275 data, and is consistent with 
mildly relativistic outflows observed in the expanding radio lobe of 
3C~84 (Asada et al. 2006). An apparent discrepancy between the model 
and data in optical-UV emission can be accommodated by remaining 
host galaxy contribution as described above.

In the standard blast wave scenario, the jet protons will contain
the majority of the jet's kinetic energy, and will be radiatively
inefficient since they are unlikely to lose their energy without
a significant observable component. If we assume they have 10
times the energy density of the electrons, the total jet power will be
$2.3\times10^{45}$ ergs s$^{-1}$, which may be inconsistent with the
estimated power required to inflate the lobe of 3C~84 against the
pressure of the hot cluster gas, (0.3 -- 1.3)$\times$10$^{44}$
erg s$^{-1}$ (Dunn \& Fabian 2004), although the
jet power in the past could be lower than at present. With the
assumption
that there is one cold proton in the flow for each radiating electron,
we will get a total jet power of $3.8\times10^{44}$ ergs s$^{-1}$, which
is consistent to within a factor of 2 of the
lobe inflation power.  Furthermore, in the context of
BL Lac and FR I unification, larger values of $\Gamma$ near the
base of the jet are expected if NGC~1275 is a misaligned BL Lac object.
Velocity gradients in the jet also help to resolve spectral
modeling issues in BL Lac objects (Georganopoulos \&
Kazanas 2003b; Ghisellini et al. 2005) and the apparent conflict
between the subluminal VLBI apparent speeds usually measured
in TeV BL Lacs (e.g. Piner et al. 2008) and the need for highly
relativistic outflows required to model their TeV emission.

A fit to the NGC 1275 data using the decelerating flow model of
Georganopoulos and Kazanas (2003b), which was developed to overcome 
these problems, is also shown in Fig.\ 7 (blue solid curve). 
In this model, the high-energy emission is due to synchrotron photons 
produced in the slower part of the flow that are 
Compton-scattered by energetic 
electrons in the faster, upstream  part  of the flow (Georganopoulos 
\& Kazanas 2003a). The  jet starts with a bulk Lorentz factor 
$\Gamma_{\rm max}=10$ and decelerates down to $\Gamma_{\rm min}=2$ over a
distance of $5 \times 10^{17}$ cm.  The cross section of the flow 
at the inlet has a diameter of 3$\times$$10^{16}$ cm, and the magnetic 
field at the base is $B=0.2 $ G. The injected power law electron 
distribution, $n(\gamma)\propto \gamma^{-p}$  has an 
index $p=1.8$, and extends from $\gamma_{\rm min}=800$ to 
$\gamma_{\rm max} = 1.0 \times 10^5$, and the particle energy density 
is higher than the magnetic field energy density by a factor of 13. 
If the protons have 10 times the energy density of the electrons, the
total jet power is $L_{\rm jet}=4.9 \times 10^{44}$ erg s$^{-1}$, 
which is still above the power needed to inflate the lobes. With an 
assumption of one proton per radiating electron, the total jet power, 
$L_{\rm jet}=6.0 \times 10^{43}$ erg s$^{-1}$, is consistent with 
this value.

The blue solid curves in Figure 7 represent the SED as 
seen at an angle $\theta=20^{\circ}$ (approximately coincident with 
$\theta_{\rm jet}$ $\sim$ 32$^{\circ}$: Asada et al. 2006).\footnote{This was 
obtained using the brightness ratio of the  northern and southern radio 
lobes, apparent velocity, and apparent  distance between the core and 
jet from recent VLBI observations.} Models with structure jets involving decelerating flows, considered
here, or a spine-sheath model  (Ghisellini et al. 2005), make predictions for 
FR I radio galaxies as potential $Fermi$ $\gamma$-ray sources.
Indeed, Ghisellini et al. (2005) 
predicted that 3C~84 would be one of the strongest $\gamma$-ray emitting radio 
galaxies above 100 MeV. 

\section{Conclusions}

We have reported the discovery that the radio galaxy 3C~84,associated
with NGC 1275 
is a source of high-energy $\gamma$ rays in the 100 MeV -- GeV range based on
data taken with the $Fermi$ Gamma ray Space Telescope between August and 
December 2008. The emission is consistent with a point source centered at the nucleus
of NGC 1275. No convincing variability is evident in the Fermi data, though there is 
a hint of a declining flux during the 
four-month observing period.  Compared with the EGRET flux upper limit, however, 
the $\gamma$-ray flux measured with $Fermi$ is almost an order-of-magnitude brighter and therefore implies that
the NGC 1275 is varying significantly on timescales from months to years. These 
results limit the amount of flux that can originate from extended galaxy-cluster 
or dark-matter--annihilation radiation to the flux upper limit measured with EGRET.

Associated with the $\gamma$-ray observations, we also report
contemporaneous and historical radio data from 3C\,84. The long-term
radio light curve appears to be brightening from an historical minimum
at 8.0 and 14.5 GHz. Core brightening during the {\it Fermi} era may be
related to its brighter $\gamma$-ray flux state than observed with
EGRET, but no unambiguous radio/$\gamma$-ray flux correlation is evident
from the historical data. 

Two jet models were used to fit the broadband SED of the nuclear emission from NGC 1275. 
A simple one-zone SSC model gives an adequate fit to the SED with a moderate Lorentz factor.
A decelerating jet model motivated by expectations of larger Lorentz factors in BL Lac/FR 1 
unification scenarios also provides a good fit to the data.

During the first year all-sky survey and beyond, we will continue to 
monitor the flux variations of NGC~1275 and 
other sources to establish the variability time scale and the fraction 
of $\gamma$-ray emission associated with compact regions. 
Future  monitoring campaigns covering
various wavebands of the electromagnetic spectrum will provide crucial data
for understanding possible correlations between high- and low-energy
bands and discriminating between models. 
The $Fermi$ $\gamma$-ray observatory will provide substantial 
insight into the physics of radio galaxies and clusters in general.

\acknowledgments

The $Fermi$ LAT Collaboration acknowledges generous support from 
NASA and DOE, the Commissariat \`a l'Energie
Atomique and the Centre National de la Recherche Scientifique/Institut
National de Physique Nucl\'eaire et de Physique des Particules in France,
the Agenzia Spaziale Italiana and the Istituto Nazionale di 
Fisica Nucleare in Italy, the Ministry of Education, Culture, Sports,
Science and Technology (MEXT), High Energy Accelerator Research
Organization (KEK) and Japan Aerospace Exploration Agency (JAXA) in
Japan, and the K.~A.~Wallenberg Foundation, the Swedish Research Council 
and the Swedish National Space Board in Sweden.
Additional support is gratefully acknowledged 
for science analysis during the operations phase from
the Istituto
Nazionale di Astrofisica in Italy and the K.~A.~Wallenberg Foundation in
Sweden, which provided a grant in support of a Royal Swedish Academy of
Sciences Research fellowship for JC.

This research made use of the NASA/IPAC Extragalactic Database (NED), 
which is operated by the Jet Propulsion Laboratory, Caltech, under
contact with the National Aeronautics and Space Administration, and 
data from the University of Michigan Radio Astronomy Observatory, which 
is supported by the National Science Foundation and by funds from the 
University of Michigan. 
The National Radio Astronomy Observatory is a
facility of the National Science Foundation operated under cooperative
agreement by Associated Universities, Inc.
%This research has made use of data from the MOJAVE database that is maintained by the MOJAVE team \citep{lister09}. 
\mbox{RATAN--600} observations are partly supported by
the Russian Foundation for Basic Research (projects 01-02-16812,
05-02-17377, 08-02-00545). The MOJAVE project is supported under National 
Foundation grant 0807860-AST and NASA-Fermi grant NNX08AV67G.

%{\it Facilities:} \facility{Fermi ()}, \facility{VLBA ()}

%\begin{table}
%\small{
%  \caption{Early LAT Observations of NGC~1275}
%\label{tab:latobs}
%  \begin{center}
%    \begin{tabular}{lccc}
%    \tableline
%    Date  & Flight Day & Primary mode & Notes\\
%    \tableline\tableline
%    2008 Jun 30-Jul 4 & 19-24 & Initial Sky Survey & Nominal Ops$^a$\\
%    2008 Jul  9-Jul 14 & 28-33 & Sky Survey & Caribratin runs$^a$\\
%    ... obs history ... & & & \\
%   \tableline
%    \end{tabular}
%   \tablenotetext{a}{Photons within 8$^{\circ}$ of NGC 1275, zenith
%   angle $z$ $le$ 105$^{\circ}$.}
%   \end{center}
%}
%\end{table}

\begin{figure}
\begin{center}
\includegraphics[angle=0,scale=0.8]{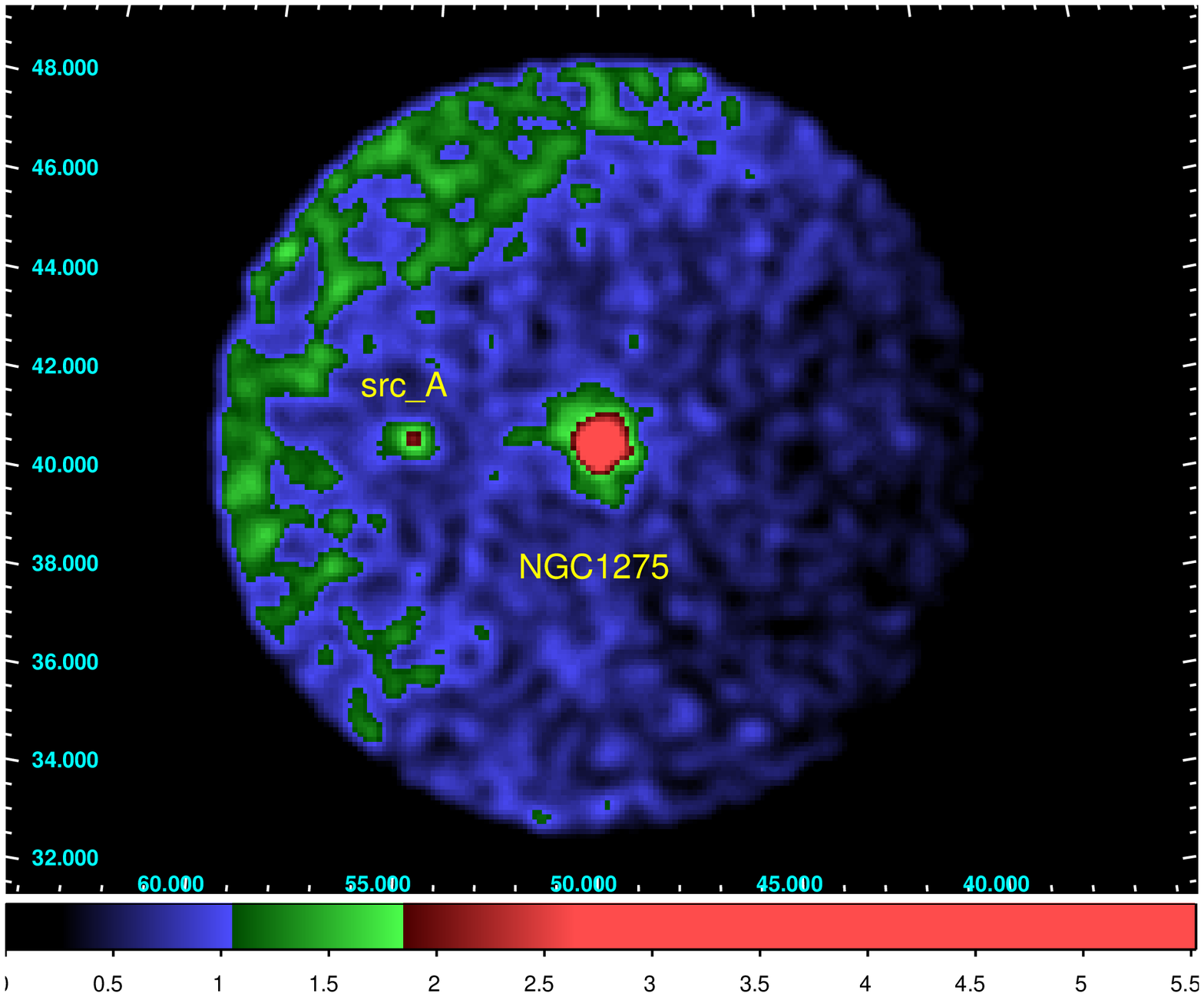}
\caption{A $\gamma$-ray sky map obtained with $Fermi$ at $E$ $>$ 200
 MeV, centered on NGC 1275 (image radius $r$ = 8$^{\circ}$, which 
is the value used throughout this paper). Sky survey data between 
August 4 and December 5 are accumulated. 
Full details are given in the text.}\label{fig:LAT_map}
\end{center}
\end{figure}

\begin{figure}
\begin{center}
\includegraphics[angle=0,scale=0.8]{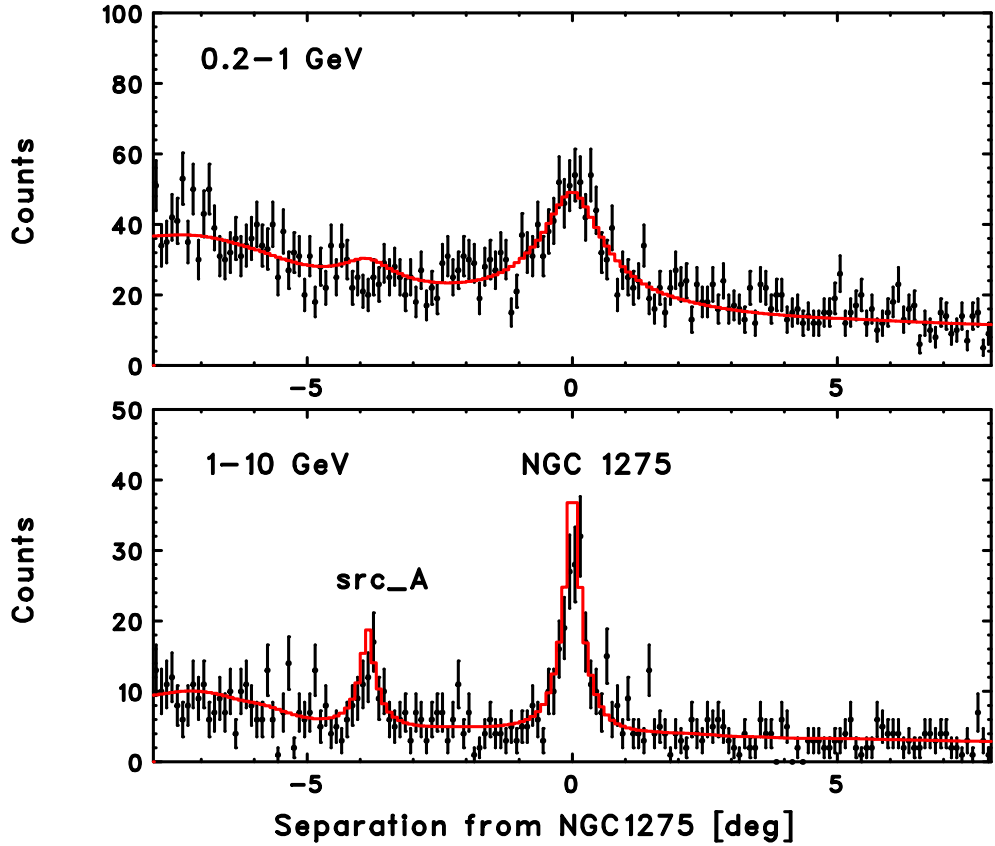}
\caption{Comparison between data (black points) and 
\textsc{gtlike} model (red solid line) in low (0.2$-$1 GeV: $upper$) 
and high (1$-$10  GeV: $lower$) energy bands for sliced projected 
count distribution. In each energy bands, $\gamma$-ray images are 
projected onto x-axis (i.e., RA plane) with sliced DEC width
 $\pm$2$^{\circ}$ centered on NGC~1275. 
Bin width is 0.1$^{\circ}$.}\label{fig:SED}
\end{center} 
\end{figure}

\begin{figure}
\begin{center}
\includegraphics[angle=0,scale=0.8]{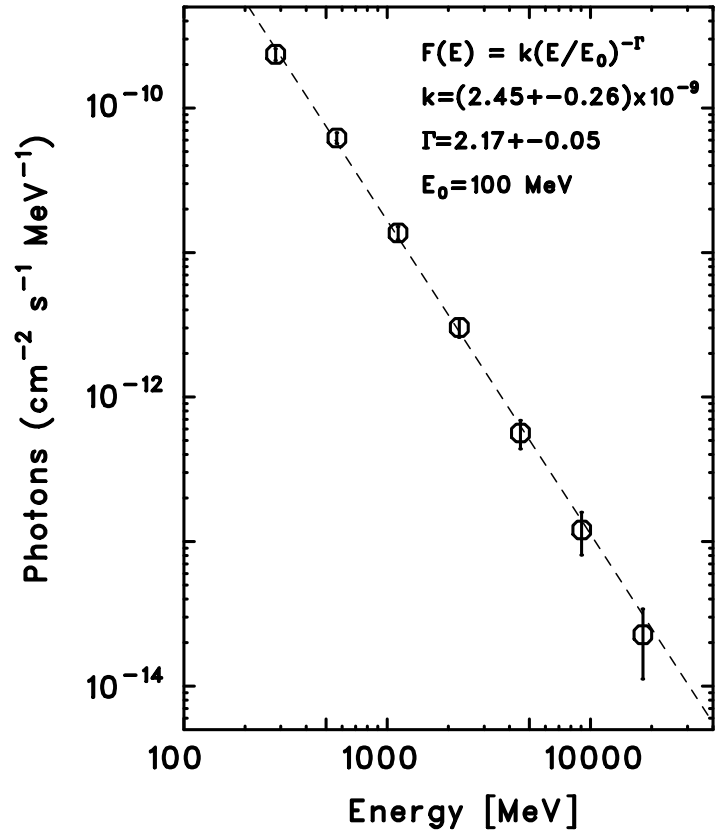}
\caption{LAT spectrum of NGC~1275 from 200 MeV to 25 GeV (open circles).
A dashed line (parameters given in the upper right of the figure) shows 
the best-fit power-law function determined from the $\textsc{gtlike}$ as 
given in the text.}\label{fig:LAT_spec}
\end{center}
\end{figure}

\begin{figure}
\begin{center}
\includegraphics[angle=0,scale=0.8]{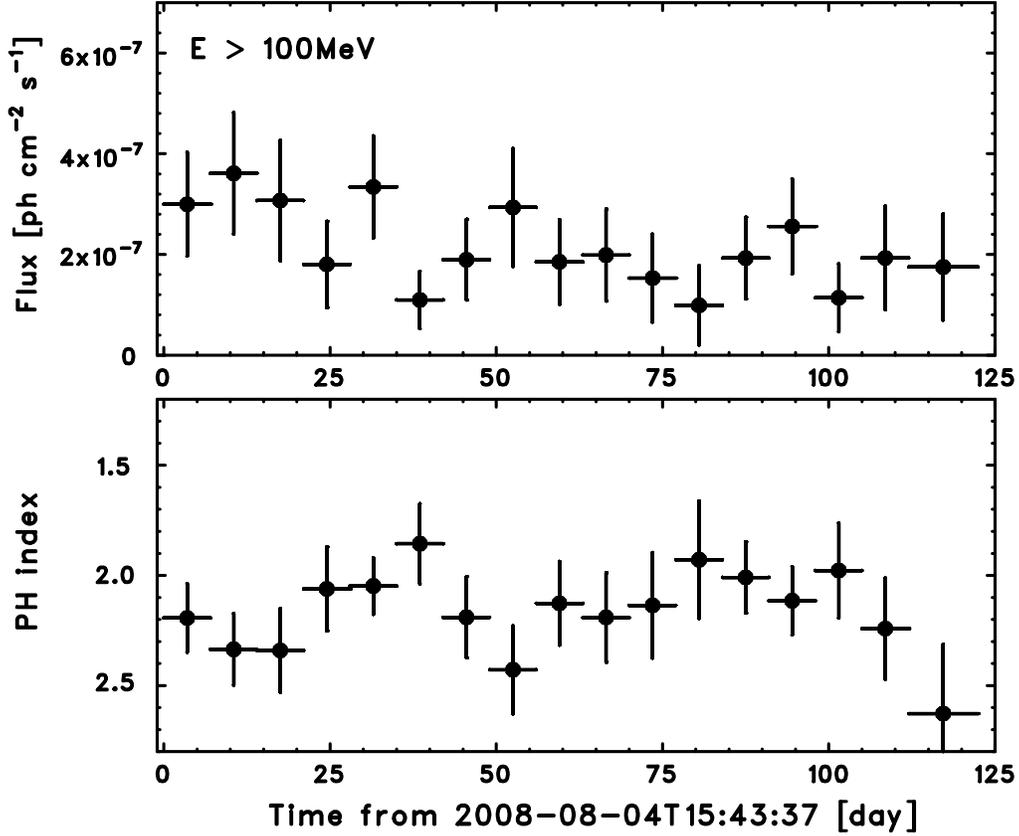}
\caption{Temporal variation of $\gamma$-ray flux and spectral index
 during the 2008 August-December observation. The observation time is
 measured from the start of the $Fermi$ observation, i.e., 2008 August 4 
 15:43:37 UT. $upper$ $panel$: changes in the  $E$ $>$ 100 MeV fluxes 
(calculated from an extrapolation of $E$ $>$ 200 MeV spectrum). $lower$ 
$panel$: changes in the power-law photon index. Background diffuse 
emission (both Galactic and extra-galactic) is fixed at the best-fit 
parameters determined from an average spectral fitting as given 
in the text, and only statistical errors are shown.
}\label{fig:LAT_LC}
\end{center}
\end{figure}

\begin{figure}
\begin{center}
\includegraphics[angle=0,scale=0.8,angle=0]{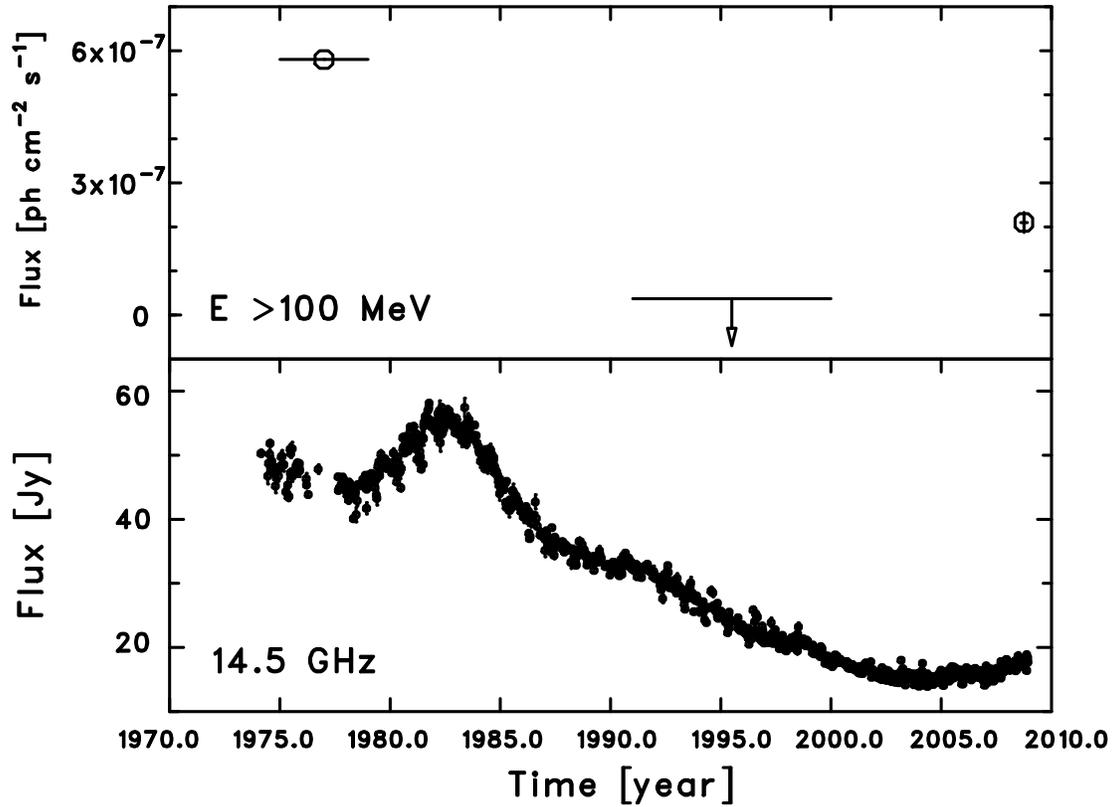}
\caption{$upper$; Historical gamma-ray activity of 3C~84 measured 
above 100 MeV. $COS$-$B$ flux in this energy range was estimated 
by assuming a differential spectral index of $\Gamma$ = 2.0. 
 $lower$; A long-term radio light curve of 3C~84 taken with the 
UMRAO at 14.5 GHz between February 1974 and December 2008. 
Data are binned with daily averages. 
The radio light curve is in a rising state during the Fermi observations.
}\label{fig:radio_img}
\end{center} 
\end{figure}

\begin{figure}
\begin{center}
\includegraphics[angle=0,scale=0.65]{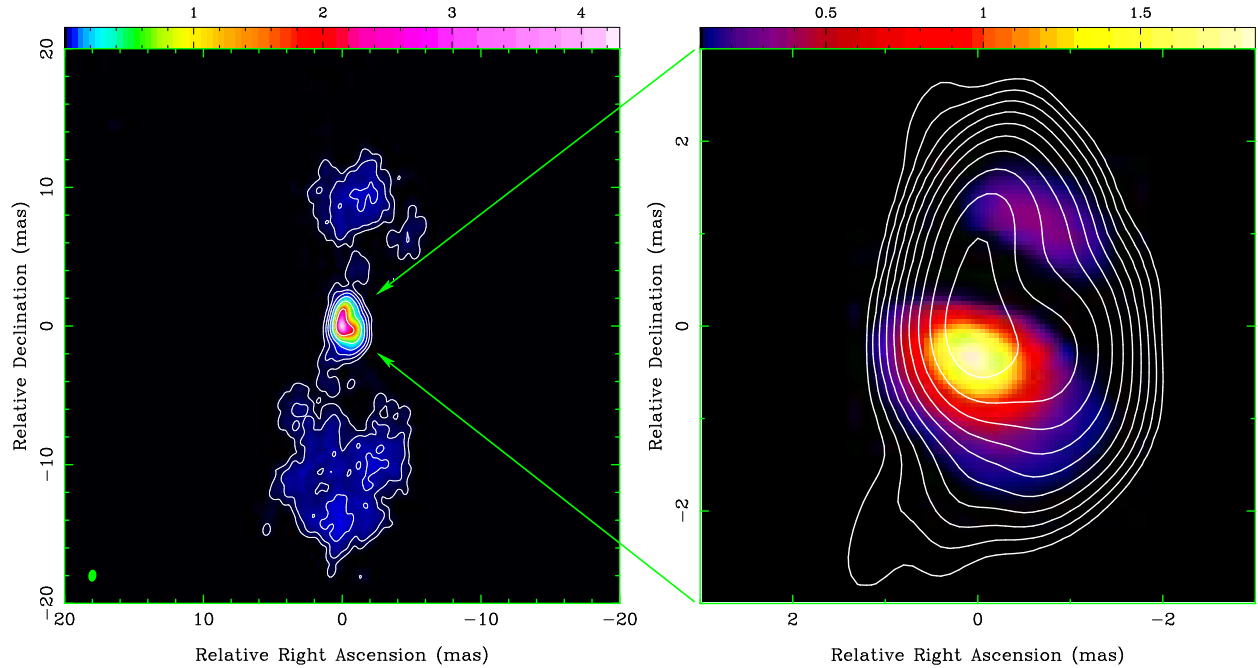}
\caption{
{\it Left:}
Naturally weighted VLBA Stokes I 15.3~GHz image of 3C\,84 observed
within the MOJAVE project (Lister et al. 2009) 
on 2008~August~25. 
It is shown both in color and in contours. The peak intensity is
4.3~Jy\,beam$^{-1}$. The beam (0.87$\times$0.58~mas)
is shown in the left corner (green).
{\it Right:}
A close~up of the central region (contours).
A difference image between 2008~August~25 and 2007~September~06 is shown
in color, where 1~mas circular restoring beam was used for both epochs.
It clearly shows that the innermost jet region has brightened
significantly, and hence a radio flare is hapenning during the {\it
Fermi} observations in 2008.
Units for color wedges are Jy\,beam$^{-1}$.
One milliarcsecond is about 0.36~pc.
}\label{fig:radio_img}
\end{center} 
\end{figure}

\begin{figure}
\begin{center}
\includegraphics[angle=0,scale=0.8]{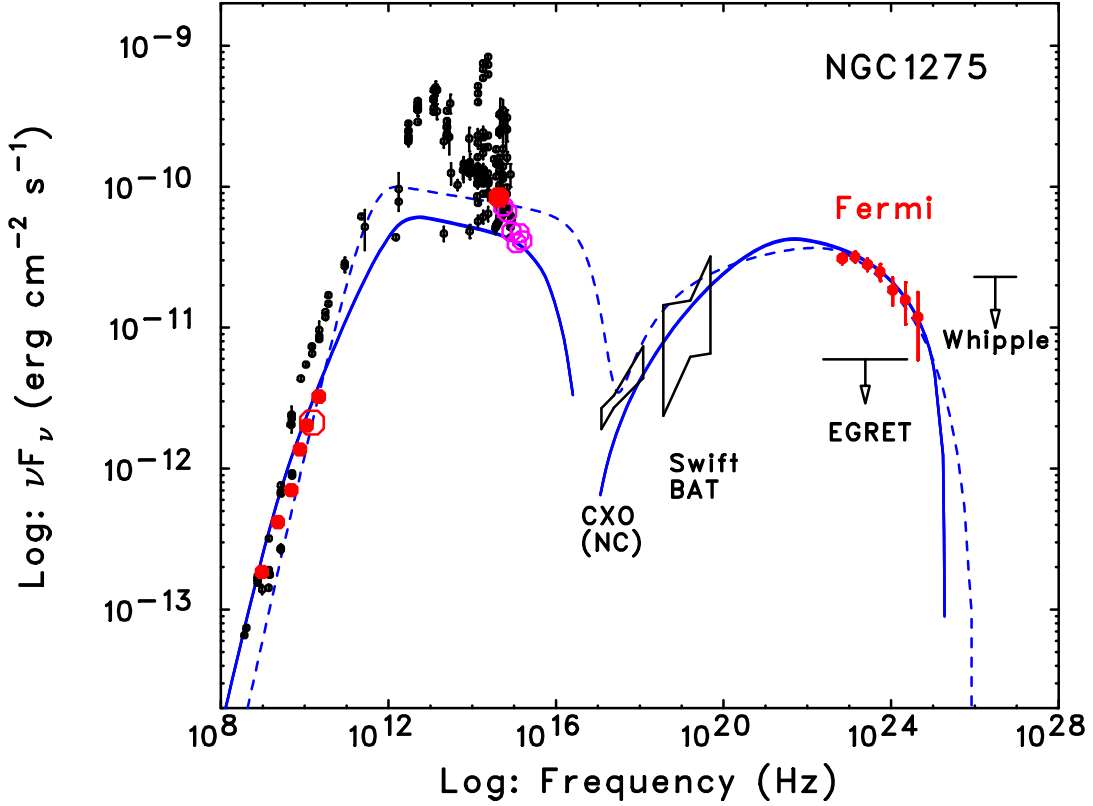}
\caption{Overall SED of NGC 1275 constructed with multiband data, using
radio (RATAN-600 in filled red circle; this work), radio core 
(MOJAVE in open red circle; this work), optical (MITSuME in red; this
 work), optical/UV (Swift/UVOT in open magenta circle; this work), 
radio to X-ray (NED), non-thermal X-ray nuclear emission 
(Balmaverde et al. 2006), hard X-ray 
(Swift/BAT; reconstructed from Ajello et al. 2009), EGRET upper limit 
(Reimer et al. 2003), Whipple upper limit 
(Perkins et al. 2006), and $Fermi$ (this work). The RATAN, MOVAVE 
and MITSuME data are contemporaneous with the Fermi data.  
Swift UVOT data come from most recent archival observation 
in December 2007. The SED is fitted with a one-zone synchrotron/SSC
model (blue dashed curve) and a decelerating flow model (Georganopoulos
 and Kazanas 2003b; blue solid curves). See text for parameters.
}\label{fig:SED}
\end{center} 
\end{figure}

\end{document}